\documentclass[prb,preprint]{revtex4-1}

\usepackage{graphicx}

\begin{document}

\title{Origin of the thermodynamic time arrow demonstrated in a realistic statistical system }

\author{Krzysztof R\c ebilas}

\address{Katedra Chemii i Fizyki, Uniwersytet Rolniczy im.\ Hugona Ko\l{}\l{}\c ataja w Krakowie.\ 
Al.\ Mickiewicza~21, 31-120 Krak\'ow, Poland}


\begin{abstract}
Emergence of one-time-direction macroscopic evolution of a classical system  of two mixed gases having different temperatures is derived and explained. 
 The analysis performed at the microscopic level, where the time-symmetric laws of mechanics govern the particles collisions, leads to a time-asymmetric macroscopic
 heat transfer equation and a theorem analogous to the Boltzmann H-theorem. Some statistical  symmetries in the velocity distribution that should be satisfied by
  the incoming and outgoing particles are pointed out. The time reversed evolution is shown to be  highly improbable  because  in this case these typical symmetries are broken. 
Additionally, some remarks explaining implicit  time-asymmetry of the Boltzmann \emph{Stosszahlansatz}  are made.

\vspace{1cm}
\noindent
Copyright (2012) American Association of Physics Teachers. This article may be downloaded for personal use only. Any other use requires prior permission of the
 author and the American Association of Physics Teachers.
The following article appeared in Am. J. Phys. 80, 700 (2012) and may be found at \url{http://ajp.aapt.org/resource/1/ajpias/v80/i8/p700_s1}.

\end{abstract}
\maketitle

\section{Introduction}
\begin{figure}
\center{\includegraphics*[width=9cm]{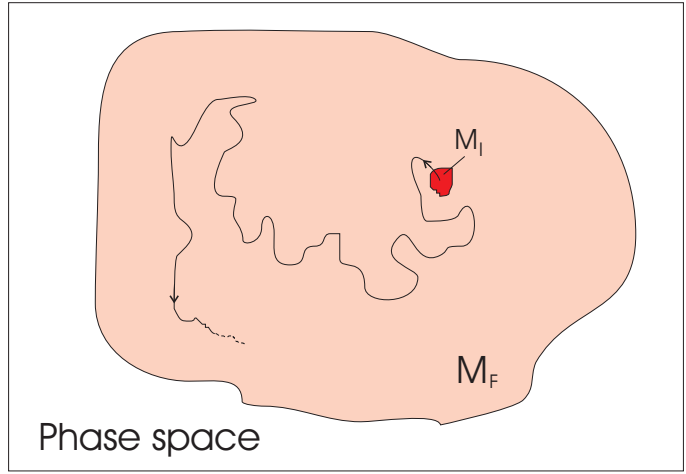}}
\caption{It is very "probable" that a microstate belonging to the macrostate $M_I$ will escape to the much more vast region of the macrostate $M_F$. To maintain the appropriate ratio of volumes of the macrostates $M_I$ and $M_F$ for typical statistical systems, the region of $M_I$ should occupy the area of $2^{10^{23}}$ smaller than the region of $M_F$, i.e. in our figure it should be many orders smaller then the size of a single electron. Thus, coming back from the macrostate $M_F$ to $M_I$ is actually very "improbable".}    
\label{Fig1}
\end{figure}

Derivation of macroscopic thermodynamic laws from microscopic dynamics is an outstanding problem of statistical physics. One of the most interesting issues is explaining the \emph{time-asymmetric} macroscopic evolution of statistical systems (the thermodynamic arrow of time).  It is well known that one-time-direction macroscopic dynamics cannot be a consequence of the \emph{time-symmetric} microscopic dynamics ruling the behavior of the microscopic constituents of a statistical system.   First arguments of statistical type explaining the macroscopic time-arrow were given by L. Boltzmann\cite{LB}:
After some constrains are removed from an isolated macroscopic system being initially in a macrostate $M_I$, the phase space volume available to the system becomes fantastically enlarged; it is then very ''probable'' that due to the random motion of particles belonging to the system, the initial microstate of the system will evolve to the newly available huge regions of phase space, most of which (almost all) pertain to a new macrostate $M_F$ (new state of equilibrium). As the new volume of phase space is enormously larger than the initial one, it is then very ''improbable'' for the system to come back spontaneously to the beginning region of the phase space, i.e. from the macrostate $M_F$ to the macrostate $M_I$ (see Fig. \ref{Fig1}).

The above-mentioned reasoning,  to some extent, counters the famous Loschmidt's \emph{Umkehreinwand} (\emph{reversibility objection})\cite{LOS} pointing out that each trajectory in the  phase space has  its precise time-reversed counterpart. Thus, preparing our system in the state $M_F$, 
there is a is a fifty-fifty chance
to find our system escaping from the state $M_F$ and going toward the state $M_I$. The answer to this objection is that the volume of
 phase space belonging to the state $M_F$ is so huge that for "overwhelmingly many" starting points of these
 time-reversed trajectories, an enormously long time is required (much exceeding the time of existence of our universe) to encounter the macrostate $M_I$ occupying a
very small volume in the phase space. Then, the transition from the macrostate $M_F$ to the macrostate $M_I$  never occurs in our world.

 Boltzmann's thoughts on the problem of the time arrow have withstood the test of time.\cite{LEB} However, the trouble is  that this kind of argumentation is based on the comparison of
 respective volumes of phase space pertaining to some macrostates. The notion of "probability" defined in terms of the volumes of phase space is valid only for the states of equilibrium 
and cannot be applied when the system is far from equilibrium, as it happens immediately after the constraints imposed on the system are released. In light of the Loschmidt's question\cite{LOS}
 such a nonequilibrium situation still remains mysterious as to why from the two time-symmetric paths of the microstate evolution (onward and backward in time) chosen is always the one that leads
 to the time-asymmetric thermodynamic behavior of the statistical system?  To understand this problem, it is necessary to trace the evolution of a microstate more directly and find some  reasons
  (of statistical type)  that apply to the nonequilibrium states and decide about their one-time direction macroscopic evolution.

The evolution of nonequilibrium states can be investigated by means of an analysis of stochastic processes for some simple models. The most famous is Boltzmann's hard
 sphere gas model leading to H-theorem\cite{LBH} which formulated in 1872 gave rise to the discussion on the problem of the arrow of time. Among the other examples, one can list 
 the Kac ring,\cite{KAC}  the "P and Q molecules" dynamics in the infinite plain,\cite{EHR} the Lorentz gas\cite{LEB1, LEB2, LEB3} and probably the best known the double-urn model invented by
 P. and T. Ehrenfests\cite{EHR2} and investigated further  by other authors.\cite{AMBEG, KAC2, METZ} For all these models, the one-time-direction behavior is obtained due to
 some reasonable probabilistic assumptions resembling the \emph{Stosszahlansatz} used for deriving H-theorem. Unfortunately, thus far it has not been clearly demonstrated
as to why these probabilistic assumptions do \emph{not} apply to the time-reversed microstates.\cite{ZEH} In other words, the \emph{Stosszahlansatz} and similar assumptions act in such a way that they select  only one type of evolution (this "normal" one) and exclude the time-reversed one. Actually, a simple explanation of how this selection comes about seems to be missing in the literature so this issue is now elucidated briefly. 

The \emph{Stosszahlansatz}-like assumptions  have the general form of master equation:
\begin{equation}
\frac{\partial \rho(s)}{\partial t}=\mathrm{gains -  losses},
\label{stos1}
\end{equation}
where $\rho(s)$ is the probability density function for particles being in the state $s$. Crucial is that:
\begin{eqnarray}
\mathrm{gains} &\propto & \rho(\neg s),\nonumber\\
\mathrm{losses} & \propto & \rho(s),
\label{stos2}
\end{eqnarray}
where $\rho(\neg s)$ is the probability density function for particles being in states other than $s$ (the symbol $\neg$ denotes the logical negation).
The losses are proportional to $\rho(s)$ because  the more particles are in the state $s$, the more of them can lose this state (in the original Boltzmann problem the particles in the state $s$ are due to collisions scattered out to another state). In turn, the less particles are in the state $s$ (then $\rho(\neg s)$ is large), the more of particles from these states different from $s$ can be transferred by means of collisions to the state $s$. Thus the gains are proportional to $\rho(\neg s)$.  As the result, the excess of particles being in $s$ is exterminated and the deficiency of such particles is augmented. This  means that $\rho(s)$ always evolves towards some equilibrium value $\rho_{eq}(s)$. The differences $|\rho_{eq}(s)-\rho(s)|$  diminish over time for any $s$, so that the entire system exhibits a monotonic evolution towards an equilibrium state. 

The \emph{Stosszahlansatz} (''the collision number assumption'') is often referred to as  ''the assumption about molecular chaos''. Actually, the molecular chaos, or the lack of any correlations between particles  is included into the \emph{Stosszahlansatz} by assuming that (\ref{stos2}) is true. The randomness of the statistical process means  that the more we have the candidates for gains (losses), the more gains (losses) actually happen. All candidates are equal and no special selection is accomplished. In other words, one can regard the validity of (\ref{stos2}) as a formal definition of the uncorrelated behavior of a statistical system.  In case of any kind of correlation, we would not be able to assume the straightforward proportionality (\ref{stos2}) because the transitions between states would happen according to some special schema following the way the particles are correlated. 

To demonstrate it  more formally and explain how the time-asymmetry is included into the \emph{Stosszahlansatz}  we  introduce a simplest possible model
where particles can assume only two states, that is $s=s_1$ or $s=s_2$.  There are only two kinds of processes proceeding at the microscopic level:  the particles change their  state from $s_1$ to $s_2$, or from $s_2$ to $s_1$. It is noteworthy that in the time-reversed evolution we observe precisely the same kinds of transitions: $s_2 \to s_1$ and $s_1 \to s_2$. However, in this case the description of the system evolution will be completely different.

\begin{figure}
\center{\includegraphics*[width=9cm]{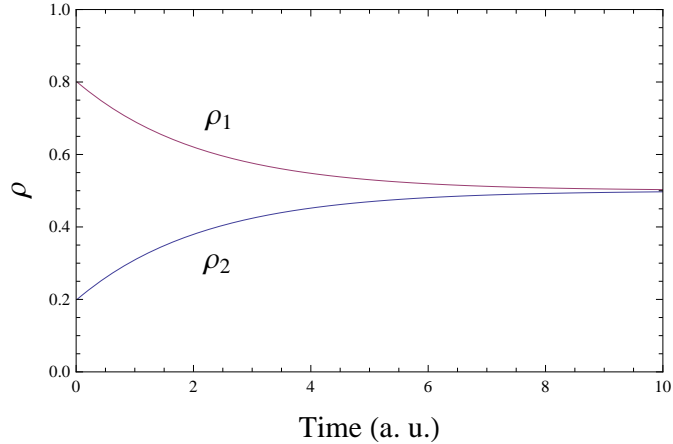}}
\caption{The time evolution of the probability density functions $\rho_1$ and $\rho_2$ determined by the \emph{Stosszahlansatz} (\ref{stos3}). }    
\label{Fig2}
\end{figure}

First the ''normal'' behavior of the system, i.e. the evolution governed by the \emph{Stosszahlansatz} is considered. Denoting $\rho_1\equiv\rho(s_1)$ and $\rho_2\equiv\rho(s_2)$, according to (\ref{stos1}) and (\ref{stos2}), we have:
\begin{eqnarray}
\frac{\partial \rho_1}{\partial t} & = & K(\rho_2-\rho_1),\nonumber\\
\frac{\partial \rho_2}{\partial t} &= & K(\rho_1-\rho_2).
\label{stos3}
\end{eqnarray}
The proportionality constant $K>0$, representing the transition rates, is assumed to be the same both for gains and losses (in the original Boltzmann reasoning, the proportionality constant depends only on the scattering cross-section and is the same both for the direct and the time-reversed collisions). The equations (\ref{stos3}) can be easily solved and the result is shown in Fig. \ref{Fig2}. 
As expected, the system evolves to an equilibrium where $\rho_1=\rho_2=\rho_{eq}$. 

If we define a function $H$ as follows:
\begin{equation}
H=\sum_{i=1}^{2}\rho_i\ln \rho_i,
\label{stos4}
\end{equation}
we find on the basis of (\ref{stos3}) that $H$ changes monotonically over time for any values of $\rho_1$ and $\rho_2$: 
\begin{equation}
\frac{\partial H}{\partial t}=K(\rho_2-\rho_1)\ln\left(\frac{\rho_1}{\rho_2}\right)\leq 0,
\label{stos5}
\end{equation}
(the  property that $\partial(\rho(1)+\rho(2))/\partial t = 0$ was used in deriving (\ref{stos5})). The inequality (\ref{stos5}) follows from the fact that if $\rho(2)-\rho(1)>0$, then $\ln(\rho(1)/\rho(2))<0$ and if $\rho(2)-\rho(1)<0$, then $\ln(\rho(1)/\rho(2))>0$. Equality holds only when $\rho(1)=\rho(2)$.
Therefore, the system exhibits the one-time directional behavior which can be expressed as the law  that $H$ never increases over time.

Let us consider the time-reversed evolution of the system. Again, we deal with the same microscopic transitions of the form $s_1 \to s_2$ and $s_2 \to s_1$ but now the relations (\ref{stos1}) and (\ref{stos2}) are time-reversed and we have:
\begin{eqnarray}
\mathrm{gains}&\propto &\rho(s),\nonumber\\
\mathrm{losses}&\propto &\rho(\neg s).
\label{stos2r}
\end{eqnarray}
or stating it explicitly for our model:
\begin{eqnarray}
\frac{\partial \rho_1}{\partial t}& = & K(\rho_1-\rho_2),\nonumber\\
\frac{\partial \rho_2}{\partial t}& = &K(\rho_2-\rho_1)
\label{stos3r}
\end{eqnarray}	
However, consequences of (\ref{stos3r}) are  very weird. We pay our attention at the evolution of the number of particles in the state $s_1$. 
Although the state $s_1$ is gained by the particles that have been previously in the state $s_2$, the number of the appropriate transitions
 $s_2 \to s_1$ (the number of gains) is by no means determined by the number of the candidates for such a transition, i.e. the particles
 in the state $s_2$.  The number of particles in the state $s_2$ appears to be completely unimportant as concerns the gains because the gains
  are now proportional to the number of particles in the state $s_1$, i.e. $\rho_1$. It is actually an extraordinary behavior. There can be very small
 number of particles in the state $s_2$ and nonetheless we can have a large number of transitions $s_2 \to s_1$ due to a large density $\rho_1$.
 It is equivalent to state that now we deal with a kind of correlation; the transitions are not random, as it was for the ''normal'' evolution where they
 were proportional to $\rho_2$; however, are mysteriously controlled in such a way to be proportional to $\rho_1$.

To learn more about discussion on the time arrow, we refer to Ref. 15, 16 and 17--19 where new ideas are investigated. A method to demonstrate the time irreversible evolution of statistical system based on a Fourier series expansion of the probability density function is presented in Ref. 20. The role played by thermal noise in the emergence of irreversible macroscopic behavior is explained on the basis of a two-dimensional model in Ref. 21. A resolution of Loschmidt's paradox for systems governed by Nos\'{e}-Newton mechanics is proposed in Ref. 22. The probability of observing Second Law violating fluctuations for non-equilibrium systems can be estimated within the framework of  Fluctuation Theorem, which is described in Ref. 23.

The aim of this paper is to provide a simple explanation of the time arrow in a realistic statistical system of two mixed
gases interacting by means of elastic collisions.  The advantage of the proposed approach is that it is not an artificial and merely heuristic model but it refers to real microscopic processes governed by the time-symmetric dynamics. The time arrow is shown for the development of an intuitive quantity such as the average kinetic energy of particles.   Moreover, the origin of the time-arrow can be easily demonstrated by referring to spatial distributions of the velocity vectors that should be satisfied for randomly moving particles before and after collisions.  These distributions appear to be significantly different (incompatible) for the incoming and outgoing particles, which is crucial for understanding the one-time direction evolution of such a system.  No \emph{Stosszahlansatz}-like assumptions are used. To formally describe the time-asymmetry, a theorem similar to the Boltzmann H-theorem is proposed.

\section{Derivation of the time arrow in a realistic system}
\begin{figure}
\center{\includegraphics*[width=9cm]{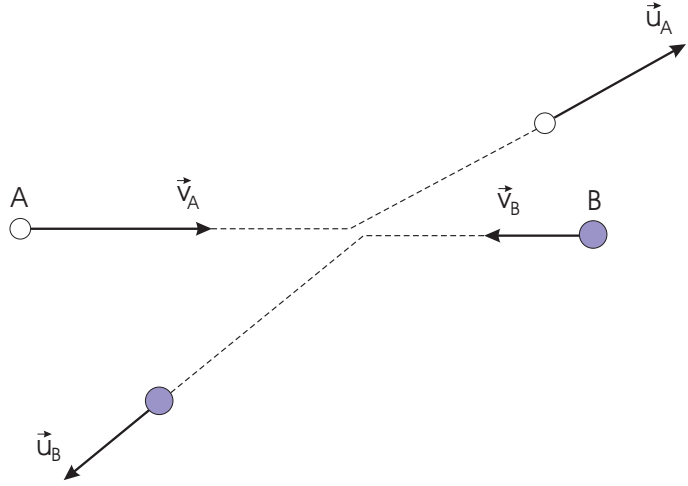}}
\caption{A collision of two particles belonging to gases $A$ and $B$. }    
\label{Fig4}
\end{figure}

Consider two diluted mixed gases $A$ and $B$ (with zero bulk velocity) initially having different temperatures $T_A$ and $T_B$ (temperature is understood as a quantity proportional to the average kinetic energy according to the relation $\langle E\rangle =(3/2)kT$). This purely theoretical initial state can be  approximately realized in practice  by removing a thermal constraint insulating the two gases from each other, each in an equilibrium state, and allowing them for mutual interactions.
Assumption is made that the particles of gases are hard-spheres
that interact by means of elastic two-particles collisions. Due to the collision of a particle $A$ with a particle $B$, the particle $A$ changes its velocity $\vec{v}_A$ it possessed just before the collision into the velocity $\vec{u}_A$; and similarly, the  initial velocity $\vec{v}_B$ of the particle $B$ changes into the velocity $\vec{u}_B$ (Fig. \ref{Fig4}):
$$
(\vec{v}_A ,\vec{v}_B)\rightarrow (\vec{u}_A, \vec{u}_B).$$

From the point of view of the system of coordinates connected with the center of mass of the two particles (CM-system), their velocities change from $\vec{v}_{A_C}$ to $\vec{u}_{A_C}$ for the particle A and from $\vec{v}_{B_C}$ to a velocity $\vec{u}_{B_C}$ for the particle B:
$$
(\vec{v}_{A_C},\vec{v}_{B_C} )\rightarrow (\vec{u}_{A_C}, \vec{u}_{B_C}).
$$

It follows from the principles of energy and momentum conservation that due to the collision the magnitude of the particle velocity remains the same  in the CM system of coordinates and only its direction is altered.\cite{KIT} So that we have:
\begin{equation}
{u}_{A_C}={v}_{A_C},\ \ \
{u}_{B_C}={v}_{B_C}.
\label{u=v}
\end{equation}
The change of the particles energy is:
\begin{eqnarray}
\delta{ E_A}= \frac{1}{2}m_A u_A^2-\frac{1}{2}m_A v_A^2\nonumber\\
\delta{ E_B}= \frac{1}{2}m_B u_B^2-\frac{1}{2}m_B v_B^2
\end{eqnarray}
and by substituting $\vec{v}_A=\vec{v}_{A_C}+\vec{V}_{CM}$,  $\vec{v}_B=\vec{v}_{B_C}+\vec{V}_{CM}$, $\vec{u}_A=\vec{u}_{A_C}+\vec{V}_{CM}$ and $\vec{u}_B=\vec{u}_{B_C}+\vec{V}_{CM}$ we get:
\begin{eqnarray}
\delta{ E_A}=  m_A{\vec{u}_{A_C}\cdot\vec{V}_{CM}}- m_A{\vec{v}_{A_C}\cdot\vec{V}_{CM}},\nonumber\\
\delta{ E_B}=  m_B{\vec{u}_{B_C}\cdot\vec{V}_{CM}}- m_B{\vec{v}_{B_C}\cdot\vec{V}_{CM}},
\label{DEK1}
\end{eqnarray}
where the property (\ref{u=v}) was taken into account.
Recalling that $\vec{v}_{A_C}=m_B(\vec{v}_A-\vec{v}_B)/(m_A+m_B)$, $\vec{v}_{B_C}=m_A(\vec{v}_B-\vec{v}_A)/(m_A+m_B)$ and $\vec{V}_{CM}=(m_A\vec{v}_A+m_B\vec{v}_B)/(m_A+m_B)$ we have:
\begin{eqnarray}
m_A {\vec{v}_{A_C}\cdot\vec{V}_{CM}}=M\left(\frac{1}{2}m_A v_A^2-\frac{1}{2}m_B v_B^2\right)-\bar{M} {\vec{v}_A\cdot\vec{v}_B}, \nonumber\\
m_B{\vec{v}_{B_C}\cdot\vec{V}_{CM}}=M\left(\frac{1}{2}m_B v_B^2-\frac{1}{2}m_A v_A^2\right)
+\bar{M}{\vec{v}_A\cdot\vec{v}_B},
\label{VACVCM}
\end{eqnarray}
where $M={2m_Am_B}/{(m_A+m_B)^2}$ and $\bar{M}=M(m_A-m_B)/2$. Eq. (\ref{VACVCM}) can be found in a slightly different form in Ref. 25.
In effect, Eqs. (\ref{DEK1}) can be written as:
\begin{eqnarray}
\delta{ E_A}=M\left(\frac{1}{2}m_B v_B^2-\frac{1}{2}m_A v_A^2\right)+\bar{M}{\vec{v}_A\cdot\vec{v}_B} + m_A{\vec{u}_{A_C}\cdot\vec{V}_{CM}},\nonumber\\
\delta{ E_B}=M\left(\frac{1}{2}m_A v_A^2-\frac{1}{2}m_B v_B^2\right)- \bar{M}{\vec{v}_A\cdot\vec{v}_B} + m_B{\vec{u}_{B_C}\cdot\vec{V}_{CM}}.
\label{DEK}
\end{eqnarray}

\begin{figure}
\center{\includegraphics*[width=9cm]{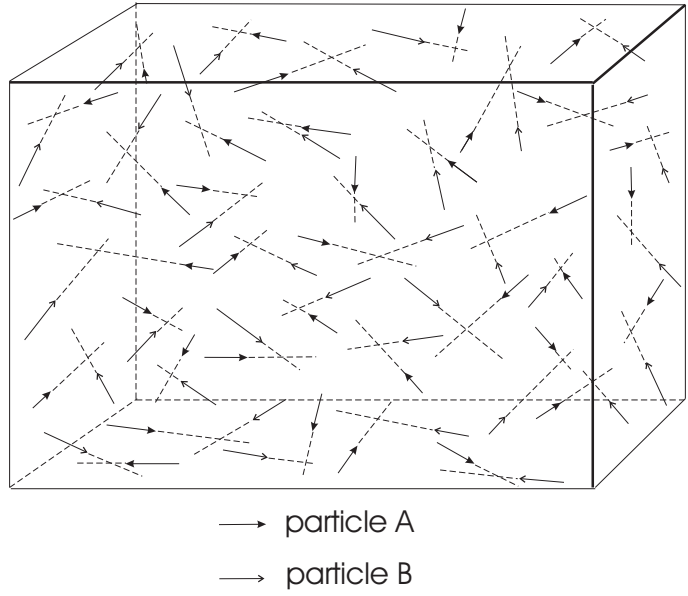}}
\caption{The symmetry $S_1(\vec{v}_A,\vec{v}_B)$: The initial velocities of particles colliding within a time $\delta t$ are uncorrelated. }    
\label{Fig5}
\end{figure}

 If we consider many collisions that proceed in the statistical system within a time interval $\delta t$, we may expect that the chaotic motion of particles makes them completely uncorrelated before
 the collisions (Fig. \ref{Fig5}),
 i.e. for the particles interacting within the time $\delta t$ we have:
\begin{equation}
S_1(\vec{v}_A,\vec{v}_B):\;\;\;\;\;\;\;\;\;\;\;\;\;\;\;\;\;\;\; \langle {\vec{v}_A\cdot\vec{v}_B}\rangle_{coll}=0.
\label{vv}
\end{equation}
The average symbol, $\langle . . . \rangle_{coll}$,  means that the average is calculated over the \emph{colliding} within $\delta t$ particles
 (not over the all particles in the gases). The property (\ref{vv}) will be referred to as the symmetry $S_1(\vec{v}_A,\vec{v}_B)$. We notice that by
 defining the temperature as a quantity proportional to the average kinetic energy we  ensure that  despite  the interacting gases are no longer in equilibrium their temperature is still well defined. 

\begin{figure}
\center{\includegraphics*[width=9cm]{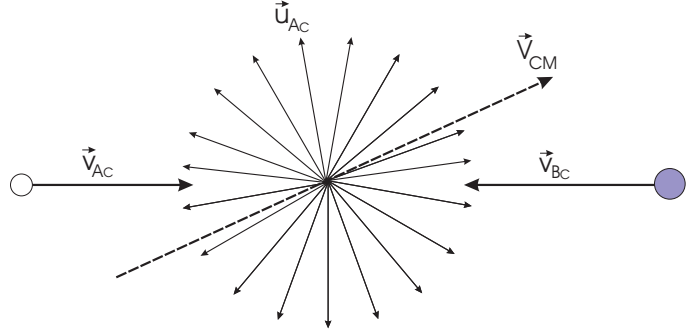}}
\caption{Visualization of the symmetry $S_2(\vec{u}_A,\vec{u}_B)$: Many particle collisions with definite velocities  $\vec{v}_{A_C}$ and $\vec{v}_{B_C}$  but random impact parameters yield the outgoing velocities $\vec{u}_{A_C}$ (and $\vec{u}_{B_C}$  -- not marked in the picture) distributed isotropically in space in terms of the CM-systems connected with the colliding pairs. }    
\label{Fig6}
\end{figure}

Now we estimate the velocity distribution after collisions and calculate $\langle {\vec{u}_{A_C}\cdot\vec{V}_{CM}}\rangle_{coll}$ and $\langle {\vec{u}_{B_C}\cdot\vec{V}_{CM}}\rangle_{coll}$. The
 theory of elastic collisions of hard-spheres demonstrates that the differential scattering cross-section in the CM-system is independent of the scattering angle. It means that if
 we consider the particles with established velocities $\vec{v}_{A_C}$, $\vec{v}_{B_C}$ and $\vec{V}_{CM}$ and differing only in the impact parameter, the outgoing velocities
 $\vec{u}_{A_C}$  and $\vec{u}_{B_C}$ will be distributed in space in CM-systems completely isotropically (Fig. \ref{Fig6}). 
 The average projection of these velocities on the velocity $\vec{V}_{CM}$  must then be zero: $\langle {\vec{u}_{A_C}\cdot\vec{V}_{CM}}\rangle_{coll}=0$ and $\langle {\vec{u}_{B_C}\cdot\vec{V}_{CM}}\rangle_{coll}=0$. The same refers to any subset 
 of colliding within $\delta t$  particles having established $\vec{v}_{A_C}$, $\vec{v}_{B_C}$ and $\vec{V}_{CM}$. Hence, we conclude that for all colliding particles holds:
\begin{equation}
S_2(\vec{u}_A,\vec{u}_B):\;\;\;\;\;\;\;\;\;\;\;\;\langle {\vec{u}_{A_C}\cdot\vec{V}_{CM}}\rangle_{coll}=0,\;\; \langle {\vec{u}_{B_C}\cdot\vec{V}_{CM}}\rangle_{coll}=0.
\label{uV}
\end{equation}
This property will be referred to as the symmetry $S_2(\vec{u}_A,\vec{u}_B)$. Now, coming back to Eq.(\ref{DEK}) we find with help of the symmetries (\ref{vv}) and (\ref{uV}) that the change of the average energy of colliding particles is given by the simple expression:
\begin{eqnarray}
\delta{\langle E_A\rangle_{coll}}=M\left(\frac{ m_B\langle v_B^2\rangle_{coll}}{2}-\frac{ m_A \langle v_A^2\rangle_{coll}}{2}\right),\nonumber\\
\delta{\langle E_B\rangle_{coll}}=M\left(\frac{ m_A \langle v_A^2\rangle_{coll}}{2}-\frac{ m_B \langle v_B^2\rangle_{coll}}{2}\right).
\label{DE}
\end{eqnarray}
Note that the equality $\langle \delta E\rangle=\delta\langle E\rangle$ was used in deriving (\ref{DE}).  If we assume\cite{REM} that that the particles colliding within $\delta t$ are representative for the whole particles of the gas $A$ and $B$
 (i.e. their average kinetic energy is the same as the average kinetic energy for the whole gas), then the average kinetic energy can
 be expressed by means of the gas temperature according to the relation $(1/2) m_A\langle  v_A^2\rangle_{coll} =(3/2)kT_A$ and $(1/2) m_B\langle v_B^2\rangle_{coll} =(3/2)kT_B$, where $k$ is the Boltzmann constant. In effect (\ref{DE}) becomes:
 \begin{eqnarray}
\delta{\langle E_A\rangle_{coll}}=\frac{3}{2}kM\left(T_B-T_A\right),\nonumber\\
\delta{\langle E_B\rangle_{coll}}=\frac{3}{2}kM\left(T_A-T_B\right).
\label{DET}
\end{eqnarray}
Note that the mutual interactions between the particles of the \emph{same} gas are unimportant because, on the basis of Eqs. (\ref{DET}), they depict no transfer of energy, thus could be neglected from the start.
The averages in Eqs. (\ref{DET}) refer to the \emph{portions} of gases that interacted within $\delta t$. To obtain the change of the average
 energy calculated with respect to the \emph{whole} gas (A or B), we introduce: $N$ -- number of all particles of a given gas ($N_A$ or $N_B$), $\delta n$ --  number of colliding within $\delta t$ pairs and  $n'=\delta n/\delta t$ - number of colliding pairs in a unit of time. Then, the rate of the change of the average energy for the entire gas is:
\begin{equation}
\frac{\delta \langle E\rangle}{\delta t}=\frac{n'}{N}\delta \langle E \rangle_{coll}
\end{equation}
and using Eq. (\ref{DET}) we obtain:
\begin{eqnarray}
\frac{\delta{\langle E_A\rangle}}{\delta t}=\frac{3}{2N_A}kMn'\left(T_B-T_A\right),\nonumber\\
\frac{\delta{\langle E_B\rangle}}{\delta t}=\frac{3}{2N_B}kMn'\left(T_A-T_B\right).
\label{DETE}
\end{eqnarray}
The above-mentioned analysis carried out on the microscopic level has led us to the macroscopic time-asymmetric \emph{heat conduction equation} (\ref{DETE}). In agreement with experiment, the transfer of energy within $\delta t$ is solely governed by the difference of the instantaneous temperatures $T_A$ and $T_B$ and  proceeds with time only in one direction: the higher-temperature gas cools down and the lower-temperature gas warms up, but not \emph{vice versa}. From the theoretical point of view, this is a very non-trivial result. We also notice that the identification of the temperature with  the average energy of particles acquires here an additional justification.  According to the experimental definition of temperature, it is a quantity that is the same for two bodies in contact and in the state of equilibrium (defined as zero flux of energy). This fact is precisely reproduced by our Eqs. (\ref{DETE}). There is no energy transfer if the temperatures \emph{understood as the average kinetic energies} are equal.

To express the achieved time-asymmetry in a concise and straightforward manner, we define an $H$ function as follows:
\begin{equation}
H=N_A\langle E_A \rangle\ln\langle E_A \rangle+N_B\langle E_B \rangle \ln \langle E_B \rangle.
\label{H}
\end{equation}
Its time derivative is given as:
\begin{eqnarray}
\frac{\delta H}{\delta t} =N_A\frac{\delta \langle E_A\rangle}{\delta t}\ln\langle E_A \rangle+N_B\frac{\delta \langle E_B\rangle}{\delta t}\ln\langle E_B \rangle,
\end{eqnarray}
where we have used the following from  Eqs. (\ref{DETE}) relation: $N_A\frac{\delta \langle E_A\rangle}{\delta t}+N_B\frac{\delta \langle E_B\rangle}{\delta t}=0$, which is, of course, the energy conservation law. Now, using explicitly Eqs. (\ref{DETE}), we get:
\begin{equation}
\frac{\delta H}{\delta t}=\frac{3}{2}kMn' \ln \left(\frac{T_A}{T_B}\right)(T_B-T_A).
\label{dH}
\end{equation}
It is evident that for any instantaneous temperatures $T_A$ and $T_B$:
\begin{equation}
\frac{\delta H}{\delta t}\leq0
\label{dHm0}
\end{equation}
and equality holds only when the gas temperatures are equal. The $H$ function changes monotonically with time and 
may be used to represent  the one-time-direction macroscopic evolution of our statistical system.  We then arrive at the
 Boltzmann type H-theorem which is  expressed here in terms of the instantaneous temperatures of gases.

\section{Origin of the time arrow}

It is intriguing as to how the time-asymmetry was introduced into our reasoning 
so that it led us to the results (\ref{DETE}) and (\ref{dHm0}). So far, we have considered time-symmetric collisions and no choice between onward and backward collisions has been  explicitly made. It appears however that
the time arrow was implicitly brought in by the statistical assumptions we named the symmetries $S_1$ and $S_2$ defined in (\ref{vv}) and
 (\ref{uV}), respectively. Namely, it should be noticed that the symmetry $S_1$ is ascribed to the \emph{incoming} velocities, $S_1(\vec{v}_A,\vec{v}_B)$, and  the symmetry $S_2$ refers to the \emph{outgoing} 
velocities, $S_2(\vec{u}_A,\vec{u}_B)$. Additionally, what is the most important, the symmetries $S_1$ and $S_2$ are \emph{incompatible} when the temperatures $T_A$ and $ T_B$ are different. This
 incompatibility (which is proved in Appendix) means that the two symmetries cannot be fulfilled at the same time by one and the same set of the velocity vectors. So that, if the initial velocities
 $\vec{v}_A$ and $\vec{v}_B $ fulfill the symmetry $S_1(\vec{v}_A,\vec{v}_B)$, they at the same time do \emph{not} undergo the symmetry $S_2(\vec{v}_A,\vec{v}_B)$,
 i.e. $\langle{\vec{v}_{A_C}\cdot\vec{V}_{CM}\rangle_{coll}}\neq 0$ and $\langle{\vec{v}_{B_C}\cdot\vec{V}_{CM}\rangle_{coll}}\neq 0$.
 Similarly, as the final velocities $\vec{u}_A$ and $\vec{u}_B $ satisfy the symmetry $S_2(\vec{u}_A,\vec{u}_B)$, they do \emph{not} exhibit the symmetry $S_1(\vec{u}_A,\vec{u}_B)$ in
 the sense that $\langle {\vec{u}_A\cdot\vec{u}_B}\rangle_{coll}\neq 0$. In brief,
\begin{equation}
S_1(\vec{v}_A,\vec{v}_B )\Leftrightarrow \neg S_2(\vec{v}_A,\vec{v}_B ),\;\;\;\;\;\; S_2(\vec{u}_A,\vec{u}_B)\Leftrightarrow \neg S_1(\vec{u}_A,\vec{u}_B).
\label{incompat}
\end{equation}

Now,  imagine that we reverse in time the just considered statistical process of energy exchange between the two gases. Then, at any time interval $\delta t$, we deal with exactly the same but time-reversed collisions processes proceeding  according to the schema:
\begin{displaymath}
(-\vec{u}_A,-\vec{u}_B) \rightarrow (-\vec{v}_A, -\vec{v}_B).
\end{displaymath}
The initial velocities have become the final ones and the originally outgoing velocities are now the incoming ones. The whole process is time reversed so that the energy
 transfer now proceeds in the opposite direction: energy flows from the gas with lower-temperature to the higher-temperature one. So the question arises: why such a process in fact never happens? 

First we must notice that the time-reversed velocities  $(-\vec{v}_A,-\vec{v}_B)$ and $(-\vec{u}_A,-\vec{u}_B)$ satisfy the same symmetries as the original velocities do because
 the \emph{time}-reversal does not influence the \emph{spatial} symmetries $S_1$ and $S_2$. It means that the symmetry $S_1(-\vec{v}_A,-\vec{v}_B)$ (i.e. that $\langle(-\vec{v}_A)\cdot(-\vec{v}_B)\rangle_{coll}=0$)   and 
 the symmetry $S_2(-\vec{u}_A,-\vec{u}_B)$, which signifies  that $\langle{(-\vec{u}_{A_C})\cdot(-\vec{V}_{CM})\rangle_{coll}}= 0,\;\;\langle{(-\vec{u}_{B_C})\cdot(-\vec{V}_{CM})\rangle_{coll}}= 0$, are obviously valid.
 It also follows that the incompatibility relations (\ref{incompat}) are correct for the time-reversed velocities as well:
\begin{equation}
S_1(-\vec{v}_A,-\vec{v}_B )\Leftrightarrow \neg S_2(-\vec{v}_A,-\vec{v}_B ),\;\;\;\;\;\; S_2(-\vec{u}_A,-\vec{u}_B)\Leftrightarrow \neg S_1(-\vec{u}_A,-\vec{u}_B).
\label{incompatrev}
\end{equation}
Therefore, in the time reversed collisions, the ''initial'' velocities $(-\vec{u}_A,-\vec{u}_B)$ 
do not fulfill  the symmetry $S_1$, i.e. they exhibit a correlation:
\begin{equation}
\neg S_1(-\vec{u}_A,-\vec{u}_B):\;\;\;\;\;\;\;\langle {(-\vec{u}_A)\cdot(-\vec{u}_B)}\rangle_{coll}\neq 0.
\label{nS1}
\end{equation}
However, such a correlation is regarded as very improbable for the incoming velocities of randomly moving particles. Here, we find the first reason why the considered time-reversed process never occurs.
Similarly, according to (\ref{incompatrev}), in the time reversed collisions the ''final'' velocities  $(-\vec{v}_A,-\vec{v}_B)$  do not realize
the symmetry $S_2$, which means that the velocities $(-\vec{v}_{A_C},-\vec{v}_{B_C})$
 are not isotropically distributed in space in CM-systems:
\begin{equation}
\neg S_2(-\vec{v}_A,-\vec{v}_B):\;\;\;\;\;\;
\langle{(-\vec{v}_{A_C})\cdot(-\vec{V}_{CM})\rangle_{coll}}\neq 0,\;\;\;\langle{(-\vec{v}_{B_C})\cdot(-\vec{V}_{CM})\rangle_{coll}}\neq 0.
\label{nS2}
\end{equation}
This however is in contradiction with the scattering cross-section property for the hard spheres collisions with random impact parameters. Again, the time-reversed collisions appear to exhibit behavior that is statistically very specific and cannot be expected for the random motion of particles.  

In summary, in the statistical system of two mixed gases, the \emph{typical} symmetry for the incoming velocities is the symmetry $S_1$ and for the outgoing ones it is the qualitatively
 different symmetry $S_2$. Such an assumption has led us to the experimentally confirmed result for the energy flow (\ref{DETE}) and the time asymmetric behavior of the $H$ function (\ref{dHm0}). 
For the time-reversed process, the roles of the symmetries would have to be exchanged: the initial one would be $S_2$ and the final one $S_1$. This, however, cannot happen in realistic system
 of randomly moving objects. If the initial symmetry was the symmetry $S_2$, it would mean that the incoming velocities are correlated (the symmetry $S_1$ broken), which is highly improbable.
 Similarly, if the final symmetry was $S_1$, it would
 be equivalent to anisotropic distribution of the outgoing velocities in CM-systems (the symmetry $S_2$ broken), which is unusual for randomly moving particles. In conclusion, we have made implicitly some important selection by assuming that the symmetries in our system change in time according to the order $S_1\to S_2$. We have discarded the (dynamically acceptable)  microstates for which the order of the symmetries is inversed. This choice, we have made for statistical reasons, has ended up  in the time-asymmetric final results (\ref{DETE}) and (\ref{dHm0}).

One may ask, what happens to the symmetry $S_2$ acquired by the particles due to collisions within the time interval $\delta t$. Because in the consecutive time interval quite a \emph{new} representative subset of
 particles is chosen for the next collisions, the \emph{memory} about the acquired earlier symmetry $S_2$ is lost. This new subset of 
particles again possesses  before the collisions the initial symmetry $S_1$ (not $S_2$). Next,   within the new time interval $\delta t$, the symmetry $S_1$ is changed 
 into the symmetry $S_2$. And so on for the subsequent intervals $\delta t_i$ and  new subsets of colliding particles:
\begin{displaymath}
S_1\shortstack[c]{{\tiny $\delta t_1$}\\$\longrightarrow$}S_2 \;\;{ {\wr}}\; \; S_1\shortstack[c]{{\tiny$\delta t_2$}\\$\longrightarrow$}S_2\;\; \wr \;\;S_1\shortstack[c]{{\tiny $\delta t_3$}\\$\longrightarrow$}S_2\;\; ...
\end{displaymath}

It is interesting to notice that in \emph{equilibrium} the symmetries $S_1$ and $S_2$ are \emph{compatible}. Actually,  if $T_A=T_B$ and 
the typical symmetries $S_1(\vec{v}_A,\vec{v}_B)$ and $S_2(\vec{u}_A,\vec{u}_B)$ are valid, then from Eqs (\ref{S1''}) and (\ref{S2''}) follows that:
\begin{eqnarray}
\langle {\vec{v}_{A_C}\cdot\vec{V}_{CM}}\rangle_{coll}=\langle {\vec{v}_A\cdot\vec{v}_B}\rangle_{coll}=0,\nonumber\\
\langle {\vec{v}_{B_C}\cdot\vec{V}_{CM}}\rangle_{coll}= \langle {\vec{v}_A\cdot\vec{v}_B}\rangle_{coll}=0,\nonumber\\
\langle {\vec{u}_{A_C}\cdot\vec{V}_{CM}}\rangle_{coll}= \langle {\vec{u}_A\cdot\vec{u}_B}\rangle_{coll}=0,\nonumber\\
\langle {\vec{u}_{B_C}\cdot\vec{V}_{CM}}\rangle_{coll}= \langle {\vec{u}_A\cdot\vec{u}_B}\rangle_{coll}=0.
\end{eqnarray}
 We observe that in equilibrium, the initial and the final velocities undergo both the symmetry
 $S_1$ \emph{and} the symmetry $S_2$. So, the roles of the incoming and the outgoing particles may be plausibly exchanged and the time-reversed
evolution is equally typical as the original one. Both these behaviors are macroscopically indistinguishable and
provide the same, equal to zero, transfer of energy between the gases.

Finally, we notice that in our reasoning, we have applied a kind of ''coarse-graining'' with respect to the time. To expect that the subset of particles colliding within $\delta t$ is representative for the entire gas, so that we can identify their average energy with the gas temperature,  and that the averages defining  the symmetries $S_1(\vec{v}_A,\vec{v}_B)$ and $S_2(\vec{u}_A,\vec{u}_B)$ are actually zero (or very close to zero), we have to chose the time interval $\delta t$ \emph{sufficiently large} to ensure we deal with
 many collisions.  If the time interval $\delta t$ was too small, the anti-thermodynamic behavior (i.e. fluctuations consisting in the heat transfer from the colder gas to the hotter one) within such a
 short time-scale would be very probable. In this case,  Eq. (\ref{DEK}) averaged over the colliding particles may give any sign of the average energy change because the
 averages $ \langle {\vec{v}_A\cdot\vec{v}_B}\rangle_{coll}$, $\langle {\vec{u}_{A_C}\cdot\vec{V}_{CM}}\rangle_{coll}$ and $ \langle {\vec{u}_{B_C}\cdot\vec{V}_{CM}} \rangle_{coll}$ are
 not necessary equal to zero and it is very probable that the average energies of the colliding particles do not represent the temperature of gasses. 

\section{Numerical simulation}
\begin{figure}
\center{\includegraphics*[width=9cm]{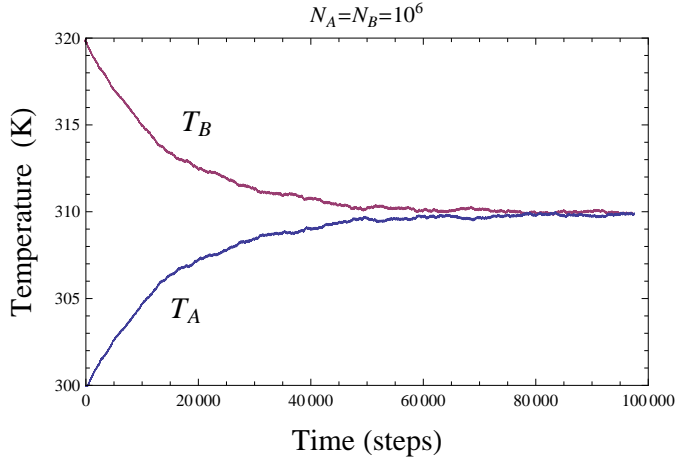}}
\caption{The time evolution of the temperature for the gases A and B obtained with the simulation,  on the basis of Eqs. (\ref{DEKaver}). }    
\label{Fig7}
\end{figure}

\begin{figure}
\center{\includegraphics*[width=9cm]{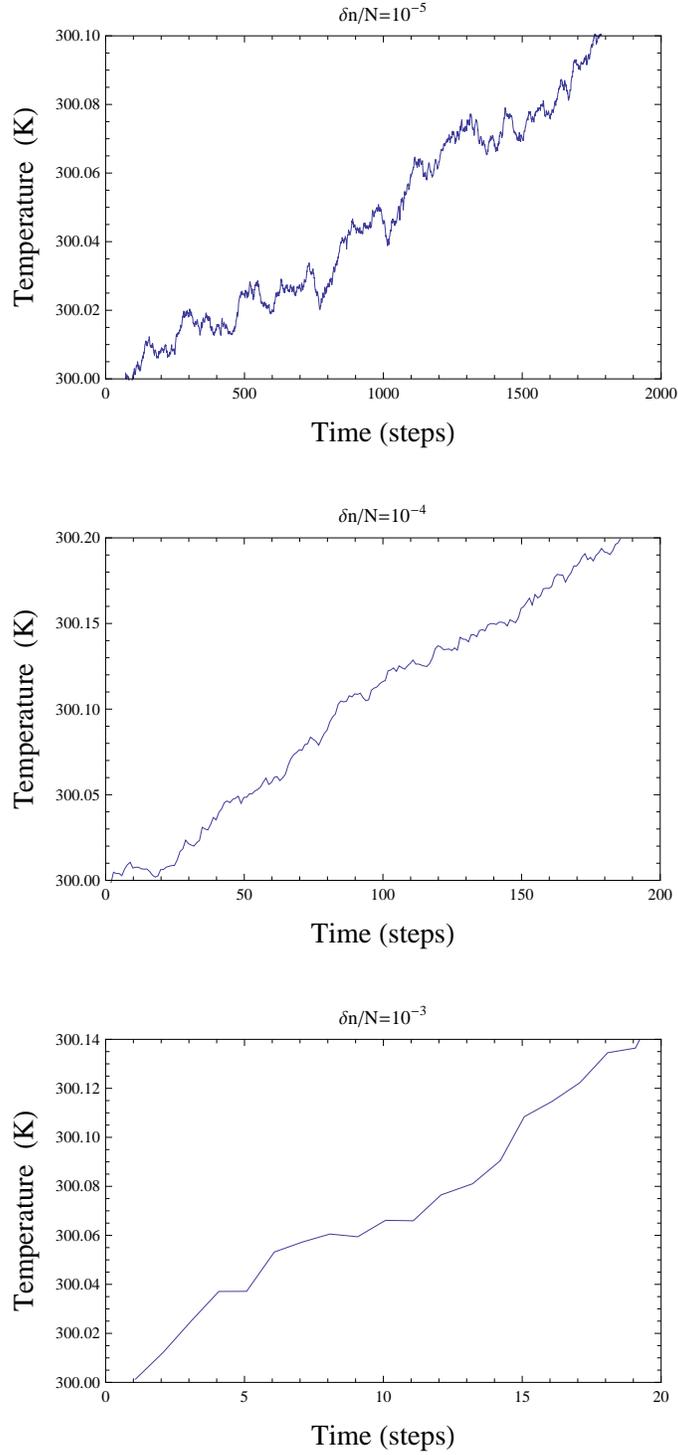}}
\caption{The enlarged plots of the time evolution of the temperature for the gas A made for different numbers of collisions $\delta n$ in a single simulation step. All plots refer to the same time period realized in different amount of steps, depending on $\delta n$.   }    
\label{Fig8}
\end{figure}

To demonstrate the behavior of the system at different time-scales, we have made a numerical simulation. Averaging Eq. (\ref{DEK}) one obtains the general expression for the change of the average energy of the entire gas A and B where no symmetries are imposed on the velocity distributions:
\begin{eqnarray}
\delta{\langle  E_A\rangle}=\frac{\delta n}{N_A}\left[M\left(\frac{1}{2}m_B \langle v_B^2\rangle_{coll}-\frac{1}{2}m_A \langle v_A^2\rangle_{coll}\right)+\bar{M}{\langle \vec{v}_A\cdot\vec{v}_B\rangle_{coll}} + m_A{\langle \vec{u}_{A_C}\cdot\vec{V}_{CM}\rangle_{coll}}\right],\nonumber\\
\delta{\langle  E_B\rangle}=\frac{\delta n}{N_B}\left[M\left(\frac{1}{2}m_A \langle v_A^2\rangle_{coll}-\frac{1}{2}m_B \langle v_B^2\rangle_{coll}\right)- \bar{M}{\langle \vec{v}_A\cdot\vec{v}_B}\rangle_{coll} + m_B{\langle \vec{u}_{B_C}\cdot\vec{V}_{CM}\rangle_{coll}}\right].
\label{DEKaver}
\end{eqnarray}
At each step of the numerical procedure $\delta n$ pairs of velocities $(\vec{v}_A,\vec{v}_B)$  are chosen at random from a  set of vectors having a definite distribution (especially a definite average magnitude) different for the gas $A$ and for the gas $B$. We have applied the Maxwell distribution. Then
 for each  velocity the magnitude of $\vec{v}_{A_C}$ and $\vec{v}_{B_C}$ is calculated  and  a direction of $\vec{u}_{A_C}=\vec{v}_{A_C}$ and $\vec{u}_{B_C}=\vec{v}_{B_C}$  is randomly
 selected. In this way we get data representing $\delta n$ collisions proceeding within a time $\delta t$ and  the respective change of the average energy can be calculated according
 to Eqs. (\ref{DEKaver}). Before the next step is accomplished, distribution of the velocities in the set from which the velocities $(\vec{v}_A,\vec{v}_B)$   are now to be selected is
 appropriately modified  by taking into account the result of the previous step that has influenced the average kinetic energy in a given gas. The simulation was made for $N=N_A=N_B=10^6$ particles. 
The ratio $m_A/m_B=1/4$ and the initial temperatures are $T_A=300$ K and $T_B=320$ K. The result showing the time dependence of the average energy (expressed in terms of temperature)
 obtained from (\ref{DEKaver})  is shown in Fig. \ref{Fig7}. 
For details showing how the evolution proceeds at different time-scales, the simulation was performed for different numbers of collisions $\delta n$ ascribed to each simulation step. We
 assume that $\delta n\sim\delta t$ so that the number $\delta n$ imitates the length of the  time-interval $\delta t$. 
To spot the difference, we magnified the plot in Fig. \ref{Fig7}. An enlarged  part of the plots for $T_A$ for different values of $\delta n$ is shown in Fig. \ref{Fig8}. 
 As can be seen, the more collisions $\delta n$ is taken in the simulation step (i.e. the longer is the time step $\delta t$), the rarer we encounter the anti-thermodynamic behavior when within  the step  the colder gas A cools down. Such an anti-thermodynamic  behavior is frequent for small samples of colliding pairs (i.e. within short $\delta t$), which reflects the fact that the symmetries $S_1(\vec{v}_A,\vec{v}_B)$ and $S_2(\vec{u}_A,\vec{u}_B)$ are not satisfied for small subsets of particles and the difference of the average kinetic energies  on the right side of  Eqs. (\ref{DEKaver}) may be arbitrary, i.e. not determined by the gas temperatures.  Therefore, the simulation confirms that the one-time-directional evolution of our system proceeds only for ''coarse-grained'' time, when the time interval $\delta t$ is regarded as sufficiently large, so that $\delta H/\delta t$ in Eq. (\ref{dHm0}) represents rather a ratio of  finite quantities than a derivative.

\section{Conclusions}
We have shown some details regarding the   evolution of microstates
 of real statistical system. At the single
 microstate level, it has been demonstrated as to why the development of the statistical system proceeds in one direction of the time. 
We have pointed out that a  \emph{statistical constraint} is imposed on typical single microstate
 and governs its evolution. This constraint is the property that the velocities of the particles {interacting} within a sufficiently large interval $\delta t$ have before collisions the
symmetry $S_1$ and the qualitatively different symmetry $S_2$  after collisions. In other words, the
 typical microscopic state of our system moves in the phase space along such a path for which the average $\langle {\vec{v}_A\cdot\vec{v}_B}\rangle_{coll}$ for the
 initial velocities and the averages
 $\langle {\vec{u}_{A_C}\cdot\vec{V}_{CM}}\rangle_{coll}$, $\langle {\vec{u}_{B_C}\cdot\vec{V}_{CM}}\rangle_{coll}$
for the final velocities give zero when calculated over the set of colliding within the time $\delta t$ particles. The time-reversed microstates are improbable because they require a breakdown of the symmetries $S_1$ and $S_2$ for the incoming and outgoing velocities, respectively.
Although for many models proposed in the literature, the development of macroscopic quantities (density, energy, etc.) toward equilibrium was clearly demonstrated, so far, to our knowledge, no general analytical determinant controlling, at any moment, the evolution of the  typical non-equilibrium microstate of realistic statistical system was provided. We believe that our work delivers some insight enabling a better understanding how the macroscopic time arrow emerges from the microscopic conditions suffered by particles. Especially, it demonstrates why the innocuous reversal of particles velocities leads to very improbable velocity distribution that is never realized in normal situations.

\section{Appendix}
To show that the symmetries $S_1$ and $S_2$ are incompatible one has to express   $\vec{v}_{A_C}$, $\vec{v}_{B_C}$   and $\vec{V}_{CM}$ by $\vec{v}_{A}$, $\vec{v}_{B}$  and gets:
\begin{eqnarray}
m_A\langle {\vec{v}_{A_C}\cdot\vec{V}_{CM}}\rangle_{coll} =\frac{3}{2}kM\left(T_A-T_B\right)-\bar{M}\langle {\vec{v}_A\cdot\vec{v}_B}\rangle_{coll},\nonumber\\
m_B\langle {\vec{v}_{B_C}\cdot\vec{V}_{CM}}\rangle_{coll} =\frac{3}{2}kM\left(T_B-T_A\right)
+\bar{M}\langle {\vec{v}_A\cdot\vec{v}_B}\rangle_{coll}.
\label{S1''}
\end{eqnarray}
It is clear that as long as $T_A\neq T_B$, if $\langle {\vec{v}_A\cdot\vec{v}_B}\rangle_{coll}=0$ (the symmetry $S_1(\vec{v}_A,\vec{v}_B )$ holds) the left side of (\ref{S1''}) cannot be equal to zero, i.e. the symmetry $S_2(\vec{v}_A,\vec{v}_B )$ is broken.

The proof for the final velocities requires a bit more effort. Using the definitions $\vec{u}_{A_C}=m_B(\vec{u}_A-\vec{u}_B)/(m_A+m_B)$, $\vec{u}_{B_C}=m_A(\vec{u}_B-\vec{u}_A)/(m_A+m_B)$ and
 $\vec{V}_{CM}=(m_A\vec{u}_A+m_B\vec{u}_B)/(m_A+m_B)$ we have: 
\begin{eqnarray}
 m_A\langle {\vec{u}_{A_C}\cdot\vec{V}_{CM}}\rangle_{coll}= M(E'_A-E'_B)-\bar{M}\langle {\vec{u}_A\cdot\vec{u}_B}\rangle_{coll},\\
 m_B\langle {\vec{u}_{B_C}\cdot\vec{V}_{CM}}\rangle_{coll}=M(E'_B-E'_A)+\bar{M}\langle{\vec{u}_A\cdot\vec{u}_B}\rangle_{coll},
\label{S21}
\end{eqnarray}
where $E'_A=1/2m_A\langle u_A^2\rangle_{coll}$ and $E'_B=1/2m_B\langle u_B^2\rangle_{coll}$.
Note that we cannot identify $E'_A$ and $E'_B$ with 
the average energy of the whole gases (the subset of particles that just collided within the  time interval $\delta t$ is no longer representative for the whole gas) and then the instantaneous temperatures cannot be introduced at this stage. But
 using Eqs (\ref{S21}) the relations (\ref{DEK}) can be written as:
\begin{eqnarray}
E'_A-E_A=M(E_B-E_A)+\bar{M}\langle{\vec{v}_A\cdot\vec{v}_B}\rangle_{coll}+ M\left(E'_A-E'_B\right)-\bar{M}\langle {\vec{u}_A\cdot\vec{u}_B}\rangle_{coll},\nonumber\\
E'_B-E_B=M(E_A-E_B)-\bar{M}\langle{\vec{v}_A\cdot\vec{v}_B}\rangle_{coll}+M\left(E'_B-E'_A\right)+\bar{M}\langle{\vec{u}_A\cdot\vec{u}_B}\rangle_{coll},
\label{S22}
\end{eqnarray}
 where $E_A=m_A\langle v_A^2/2\rangle=3/2kT_A$ and $E_B=m_B\langle v_B^2/2\rangle=3/2kT_B$. After simple 
calculations we get from the last equations that:
\begin{equation}
E'_A-E'_B=E_A-E_B-\frac{2\bar{M}}{1-2M}\langle{\vec{u}_A\cdot\vec{u}_B}\rangle_{coll}+\frac{2\bar{M}}{1-2M}\langle{\vec{v}_A\cdot\vec{v}_B}\rangle_{coll}.
\label{posr}
\end{equation}
Inserting it into Eqs (\ref{S21}) we get:
\begin{equation}
\begin{array}{l@{\vspace{0.3 cm}}}

m_A\langle {\vec{u}_{A_C}\cdot\vec{V}_{CM}}\rangle_{coll}= \frac{3}{2}kM\left(T_A-T_B\right)+\frac{2M\bar{M}}{1-2M}\langle {\vec{v}_A\cdot\vec{v}_B}\rangle_{coll}-\frac{\bar{M}}{1-2M}\langle{\vec{u}_A\cdot\vec{u}_B}\rangle_{coll},\\
m_B\langle {\vec{u}_{B_C}\cdot\vec{V}_{CM}}\rangle_{coll}= \frac{3}{2}kM\left(T_B-T_A\right)-\frac{2M\bar{M}}{1-2M}\langle {\vec{v}_A\cdot\vec{v}_B}\rangle_{coll}+\frac{\bar{M}}{1-2M}\langle{\vec{u}_A\cdot\vec{u}_B}\rangle_{coll},
\label{S2''}
\end{array}
\end{equation}
It is clear that if $T_A\neq T_B$ and the symmetries $S_1(\vec{v}_A,\vec{v}_B)$ and 
$S_2(\vec{u}_{A_C},\vec{u}_{B_C})$ are valid, then $\langle{\vec{u}_A\cdot\vec{u}_B}\rangle_{coll}\neq 0$, i.e. the symmetry $S_1(\vec{u}_A,\vec{u}_B)$ is broken.

\section*{Acknowledgments}
I wish to thank P. Prawda for inspiration to elaborate the problem of the thermodynamic time arrow, which resulted in this presentation. I am grateful to the anonymous reviewers for their constructive remarks and suggestions that allowed us to present some issues more clearly and adequately.

\end{document}